# Studying the nucleus of comet 9P/Tempel 1 using the structure of the Deep Impact ejecta cloud at the early stages of its development

Ludmilla Kolokolova[a], Lev Nagdimunov[a], Michael A'Hearn[a], Ashley King[b], Michael Wolff[c]

[a]University of Maryland, College Park, MD 20742, USA,
[b]Stanford University, 452 Lomita Mall, Stanford, CA 94305, USA,
[c]Space Science Institute, 4750 Walnut Street, Boulder, CO 80301, USA

*Corresponding author:*
Tel.: +1-301-405-1539
Fax.: +1-301-405-3538
*Email address: Ludmilla@astro.umd.edu (L. Kolokolova).*

**Abstract**
The paper presents an attempt to extract information about the comet 9P/Tempel 1 nucleus from the characteristics of the ejecta cloud produced by the impactor of the Deep Impact mission. For this purpose we use two techniques. We first study the shadow cast on the nucleus surface by the ejecta cloud and investigate how areas of different brightness are related to the varying optical thickness or albedo of the ejecta cloud. The shadow was seen during the first 2.0 seconds after the impact (afterward it became obscured by the ejecta cloud). We have found that all brightness variations in the shadow are the result of the surface inhomogeneities, indicating that during first 2.0 seconds the ejecta cloud was homogeneous within the MRI spatial resolution. Our second technique is to study the obscuration of the nucleus limb by the ejecta. This study covers the period 0.76- 68.8 seconds after impact and is based on comparison of the ejecta cloud brightness on the limb and just beyond the limb. At this stage we do see inhomogeneities in the ejecta cloud that relate to the albedo and optical thickness variations in the ejected dust. Specifically, we have found two distinct bands of low optical thickness and one band of a high optical thickness. Based on crater formation ideas we estimate the depth of excavation of the ejected material for the found inhomogeneities and, thus, define a potential layering structure for the comet nucleus, Our estimates suggest that the low-optical thickness material was excavated from a depth of 15 - 18 and 30 – 32 meters in the case the porous nucleus material and 37 - 46 and 87 - 93 meters in the case of a non-porous nucleus material, and a layer of high optical thickness originated from the depth 9-11 m for porous material or 20-23 m for non-porous material. Based on the crater diameter estimates, we expect that the real depth of the layers is between these two cases. The rest of the ejecta do not show any signs of layering but have significant azimuthal inhomogeneity with clumps of high optical thickness and patterns of high albedo.

1. Introduction

The study of pristine materials from the comet interior was the main goal of the NASA space mission Deep Impact (A'Hearn et al. 2005a, 2005b). Hundreds of images, obtained with a high temporal and spatial resolution by the Deep Impact High and Medium Resolution Instruments, showed how the cloud of the materials, excavated from the nucleus interior, was developing and how its structure was changing with the time from impact. The amount of the ejected dust and its change as the ejecta cloud was forming can provide information about the properties of the nucleus material and the nucleus structure. For this reason, the Deep Impact (hereafter DI) ground-based campaign included significant efforts to observe the DI ejecta cloud. The cloud was first spatially resolved 20 minutes after impact by Earth-orbiting telescopes at visible and UV wavelengths (Meech et al. 2005). Later ground-based telescopes worldwide imaged the expanding cloud of dust and gas in the visible and infrared. The majority of the ground-based data were obtained hours after impact. At this time, the difference in the velocities of the dust particles mixed together the grains from different depths of the nucleus. Additionally, the dust had already been altered by the solar radiation. Thus, there are no other data than in situ images and spectra taken at early stages of the ejecta cloud development that allow us to see the most unchanged and unsorted materials in their sequential appearance in the cloud from different depth of the nucleus. The images of the cloud in the first minutes or even seconds after the impact and the change in the cloud's brightness with time can provide a great source of information not only about the spatial distribution and temporal changes in ejecta grains but also the properties of fresh "parent" materials from the comet nucleus. In this paper we attempt to extract properties of the ejecta grains using a sequence of in-situ images and by combining pre- and post-impact images.

We will use two main approaches to analyze the data. Both of them measure the optical thickness of the cloud and its change with time to allow us to obtain some idea about number density and albedo of the dust particles. One of the approaches uses pre- and post-impact images of the limb of the nucleus to determine its obscuration produced by the ejecta, providing the optical depth of the cloud. The other approach uses the shadow that the optically thick part of the cloud casts on the nucleus. These reveal not only properties of the pristine comet materials but also their temporal and spatial variations, leading to a better understanding of the changes in comet material with the depth, i.e. shedding light on the structure of comet nuclei. This study allows an examination of the validity of the hypothesis about layered structure of comet nuclei (Belton et al. 2007). This, in turn, helps us to understand the formation and evolution of Jupiter Family comets, and improves our knowledge about the processes in protoplanetary nebulae.

## 2. Observations and their analysis

To study the change in the dust properties of the ejecta cloud with time we use images taken by the Deep Impact Medium Resolution Instrument (MRI) that are available through the NASA Planetary Data System, PDS (McLaughlin et al. 2014a). The images taken with the MRI are available for many seconds before and after the impact, having a time resolution of 65 msec just after impact with gradually greater spacing after impact as the ejecta are moving more slowly. In this paper we study the data taken just before the impact and during the first 68.8 seconds after the impact, thus limiting the change in geometry due to motion of the flyby spacecraft. The instrument had a variety of filters (the filter wheel contained two clear apertures and eight other filters, among them three narrow-band gas filters and the following dust filters: 345/6.8 nm, 526/5.6 nm, 750/100 nm, 950/100 nm; the numbers show the central wavelength and the bandpass of the filters, see details in Hampton et al. 2005). However, we concentrate on the data taken with the clear filter 650/ 700 nm as it provides the best signal-to-noise values. For our study of the early stages of the ejecta development we will not worry about the gas contamination: we have checked the data obtained with the gas filters and concluded that within the considered time period the gas contamination was low. In some cases, specified below, we also use selected Deep Impact High Resolution Instrument (HRI) images (McLaughlin et al. 2014b).

**2.1 First two seconds after the impact: study of the ejecta shadow.**

We measured the optical depth of the ejecta cloud and its spatial and temporal changes using the shadow that the optically thick part of the cloud casts on the surface of the nucleus. A detailed description of the shadow and its evolution with time can be found in Schultz et al. (2007) and Richardson et al. (2007). Using images of the type shown in Fig. 1, we calculated the optical depth of the cloud, comparing the brightness of the same part of the nucleus before the impact and after, when it is in the shadow. The loss in the brightness of the surface due to shadowing allows us to see how much light was blocked by the cloud, thus obtaining the optical depth of the ejecta. A detailed radiative-transfer study of the ejecta cloud and its shadow was presented in Nagdimunov et al. (2014). Using the package Hyperion (Robitalle 2011) we modeled one of the HRI images (hv9000910_007, acquired 1.036 s after impact) trying to reproduce the correct brightness of the ejecta cloud and the nucleus, including the shadowed area. The result of this modeling allowed us to estimate the number density of the dust particles ($1.5 \times 10^4$ particles/$cm^3$), their size distribution (power law with power 3 ranging particles of radius 0.1-100 micron) and a dust/ice mass ratio that appeared to be 1-1.6 for the case of the individual particle density of the dust material 0.4 $g/cm^3$ and 3.6-21.6 for the case of the individual particle density 1.75 $g/cm^3$.

The next step in our study was an analysis of the variations in the shadow brightness distribution, which we believed indicated variations in the internal structure of the ejecta cloud. These variations, in turn, could be either variations of the optical depth or the albedo of particles in the ejecta plume and could provide information about the cratering process as well as properties of the outer layers of the nucleus.

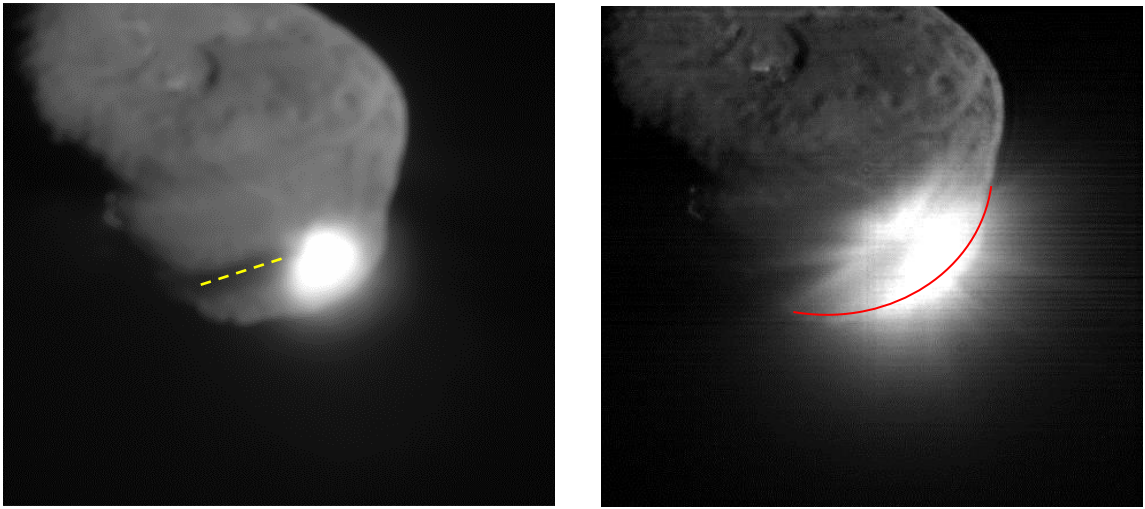

**Figure 1.** Left: HRI Image of the Tempel 1 nucleus with a shadow cast by the impact ejecta at time 1.036 s after impact. The dashed line shows the position of the scan we took through the shadow axis. Right: HRI image of the ejecta 13 s after impact. The line shows approximate position of the nucleus limb. The brightness of the ejecta on the limb and right beyond the limb is used in Section 2.2 to estimate the optical depth of the ejecta.

From an initial analysis of the MRI post-impact images (we used the images from mv9000910_074 to mv9000910_96), we found an inhomogeneous structure of the shadow and assumed that this indicates inhomogeneity in the optical depth of the ejecta plume, i.e. inhomogeneities in concentration and/or albedo of the dust particles. We assumed that this indicated inhomogeneity in the excavated material - a layering of the comet nucleus unfolding with excavation time. To see if these inhomogeneities behave as would be expected for sequentially excavated layers of the nucleus (i.e. moving up along the ejecta cone with time) we examined a sequence of MRI images in the time range 0.5 - 2.0 s and took a scan through the center of the shadow from the impact site and outward to the terminator (Fig. 1, left). Note that we could not use the images for time smaller than 0.5 s after the impact as at that time the shadow was very small and covered by the ejecta cloud. To remove the influence of the nucleus albedo variations, we made the scan through the same area on the pre-impact image of the nucleus and calculated the ratio of the brightness taken from post-impact image to the pre-impact image. At each point of the scan, such a ratio is a ratio $A*(I-I_s)$ to $A*I$, where $A$ is albedo of the given point on the surface, $I$ is the intensity of the solar radiation, and $I_s$ is the intensity removed from the solar radiation due to absorption and scattering by the ejecta particles. It is evident that considering the ratio we remove effect of the surface albedo.

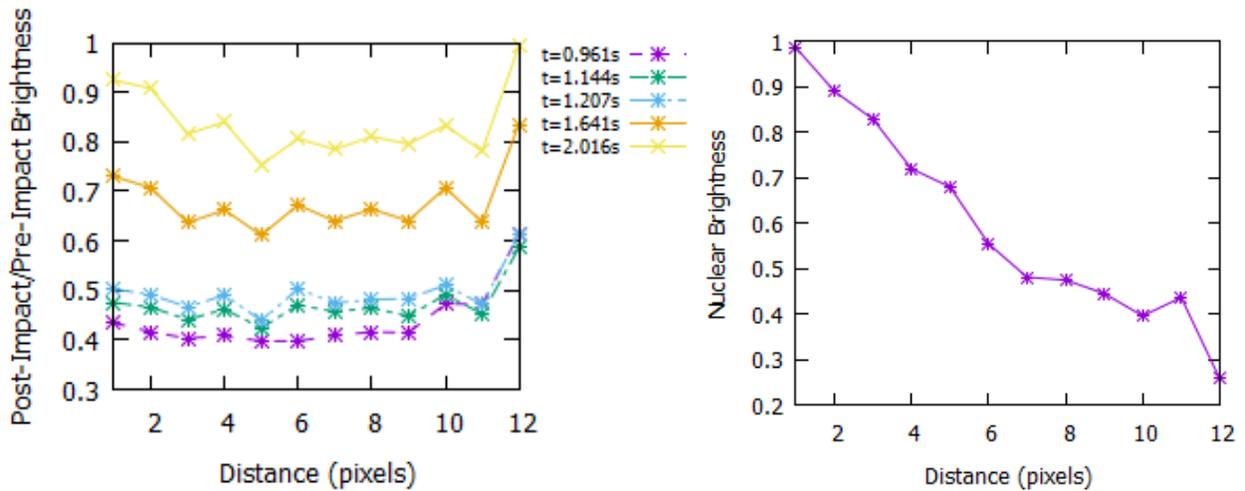

**Figure 2.** Left: Scans through the axis of the ejecta shadow for different times after the impact, MRI data prepared as the ratio of post- to pre-impact images. Right: Nucleus (pre-impact image) scan for MRI (t = -0.031s); units are $W/m^2/sr/\mu m$. Times are relative to the moment of impact. Note that the brightness variations in the left and right panels are located at the same positions; the opposite behavior of the brightness features (minima at the left figure correspond to maxima at the right features) results from dividing homogeneous post-impact images to inhomogeneous pre-impact images.

Figure 2 shows the scans of the ratio of post-impact to pre-impact image along the axis of the shadow. One can see several brightness features (pixels 3, 5, and 11) in the scans that may be caused by more optically thick parts of the ejecta cloud or be filled with darker dust particles. However, all the features do not move with time, making it doubtful that they are formed by inhomogeneities in the ejecta material as inhomogeneities, formed by variations of optical depth or particle albedo in the excavated materials, should be seen at larger distances for later time. Their stable position on the scans suggests that they are not associated with the internal structure of the expanding ejecta cloud. To investigate the cause of these features, we took scans of the same area before the impact. In the right panel of Fig.2 we show the pre-impact scans for MRI images. We can see that the brightness variations in the shadow coincide with the brightness variations of the nucleus itself. Thus, the shadow inhomogeneity reproduces a nucleus inhomogeneity. Notice that they are not albedo variations, which, as we mentioned above, should be canceled as we consider the ratio of post- and pre-impact images. Likely, they are related to the scattering of light on some structural features of the surface, e.g. crater walls.

From this we conclude that the ejecta cloud itself was rather homogeneous during the 0.5-2.0 s analysis interval, indicating homogeneity of the layer of the nucleus that was excavated during that time. We estimate the depth of the layer excavated during the first 2.0 s in Section 3.

## 2.2 Development of the ejecta up to 68.8 seconds after impact.

The optical depth of the cloud can be also detected through the obscuration of the limb by the ejecta that can clearly be seen in Figure 1 (right). We used this approach for studying the ejecta cloud during the period 0.76 - 68.8 s after (there are no reliable data on the ejecta crossing the limb for times earlier than 0.76 s after impact). Using an image taken just before the impact, at each point on the downrange limb of the nucleus, we measured the brightness of the nucleus limb above the background, $I_n$. Then on the images after the impact we measured the brightness at the same locations on the nucleus limb, $I_l$, and immediately beyond the limb, $I_b$. We assume that the dust that produces obscuration on the limb is very similar to the dust just outside the limb at the same location on the limb (the same azimuth) and moment of time, and therefore that the brightness of the ejecta just beyond the limb may be used to represent the brightness of the dust on the limb. To be more precise, we carried out a polynomial extrapolation of the out-of-nucleus brightness to the limb to get second order corrections of the brightness of the ejecta on the limb. The brightness on the limb after impact is produced by the light scattered by the ejecta dust particles and the light scattered by the nucleus and attenuated by the ejecta cloud. The latter is equal to $I_l - I_b$ and is also equal to $I_n e^{-\tau}$ that allows us to determine $\tau$, the optical depth of the ejecta. The obscuration was determined on the limb of the nucleus for a variety of the points along the nucleus perimeter, i.e. for azimuths from 160° to 290° and for a variety of the times after impact up to 68.8 seconds thus sequentially covering new strata of the ejecta cloud that cross the limb. The azimuth was measured along the nucleus perimeter counterclockwise with the zero azimuth pointing to the top of the image (see Fig. 1 right). Note that as the ejecta plume expanded, its projection on the nucleus covered a larger part of the nucleus perimeter, thus, increasing the range of azimuths.

In the case of small optical depth, for an ensemble of particles, the average single-scattering albedo of the particles, $\omega = \sigma_{sca} / \sigma_{ext}$, can be calculated as the ratio of the intensity of the light scattered by the dust to the intensity of the light lost due to the extinction by the dust particles. As mentioned above, we assume that the properties of the cloud are the same on the limb and immediately beyond the limb. This means that the geometrical thickness of the cloud, the number density of the ejecta grains, and their optical properties defined by their size and composition are the same in both these almost identical locations. These grains produce the optical depth τ extracted from the images as described above that by definition is τ = $n·L·\sigma_{ext}$, where $n$ is the number density of the particles, $L$ is the geometrical thickness of the cloud. The same grains scatter light producing the brightness beyond the limb, $I_b$, that can be defined as $I_{sun}·n·L·\sigma_{sca(62.9°)}$ where $\sigma_{sca(62.9°)}$ is the scattering cross-section at the phase angle 62.9º (the phase angle of the MRI observations) and $I_{sun}$ is the intensity of the sunlight at the heliocentric distance of the comet. One can see that from the expressions for optical depth and brightness it is possible to determine the single-scattering albedo as $\omega = I_b/(I_{sun}·\tau)$. Note, that this is not a traditional single-scattering albedo but the albedo at the phase angle of the observations that was equal to 62.9°. This means that the obtained values of albedo are smaller than the values we would obtain if we calculated the traditional single-scattering albedo of the ejecta grains. Where the optical depth is high, we use a 1D radiative transfer approach similar to the one described in Kokhanovsky (2004), Chapter 3, to estimate the albedo. We use the Henyey-Greenstein function as the phase function of the ejecta grains and calculate the optical depth counting on the Sun light scattered/absorbed by the cloud and the one that was transmitted through the cloud, reflected from the surface of the nucleus (whose reflectivity we know through the before-impact brightness of the nucleus, $I_n$) and then passed through the cloud again. Calculations are done for a variety of values for the Henyey-Greenstein parameter $g$ and albedo of the particles. The best fit parameters for the optically thick parts of the cloud demonstrate low values of albedo that stay in the range 0.02 - 0.25 and a rather stable value of g that varied only within the range 0.68-0. 70.

Figures 3 shows the results of the analysis described above. The left panel of Figure 3 shows results for the time period 0.76 -13 s (vertical axis of the plots) and the range of azimuths from the impact site described above (horizontal axis) and its right panel shows the same data but for the time range 0.76 - 68.8 s. The brightness of the ejecta on the limb is shown on the top of Fig. 3, the optical depth calculated as described above is shown in the middle, and the single scattering albedo is shown on the bottom of Fig. 3. Note that, in accordance with the results of **Section 2.1,** the optical depth of the ejecta is quite homogeneous for time earlier than two seconds after impact. However, there are noticeable bands of low optical depth at 8 - 13 s and 60 - 68.8 s. Although the 60 - 68.8 s band looks rather natural as the values of the optical depth already show some decrease in approaching the band, the 8 - 13 s band begins with a dramatic drop in the optical depth, suggesting a potential artifact. However, a careful analysis of the data showed that this is an artifact of the particular display in Figure 3 that corresponds to a large reduction in the frequency of the MRI sampling that reduced from ~0.11 s to 0.75 s. As a result, the data shown for times after 8 s have a large time gap between them that exaggerates the abruptness. Our analysis of the data showed that the optical depth had already began decreasing in many azimuths at time prior to the visible sharp cutoff and the smoothness of this decrease is disrupted by the big time gap in the data. Notice also that there are no suspicious features in the brightness for the time after 8 sec, the band in the ejecta cloud is seen only when the optical depth is calculated. Consequently we conclude that the band of low optical depth at 8-13 seconds is most likely real. It can be also seen that both bands of low optical depth are associated with a higher single-scattering albedo, although the albedo inhomogeneity does not produce any band-like feature and is concentrated within the azimuths 240°-270° for both bands. Both albedo and optical depth show numerous azimuthal inhomogeneities that are rather randomly distributed within the ejecta cloud.

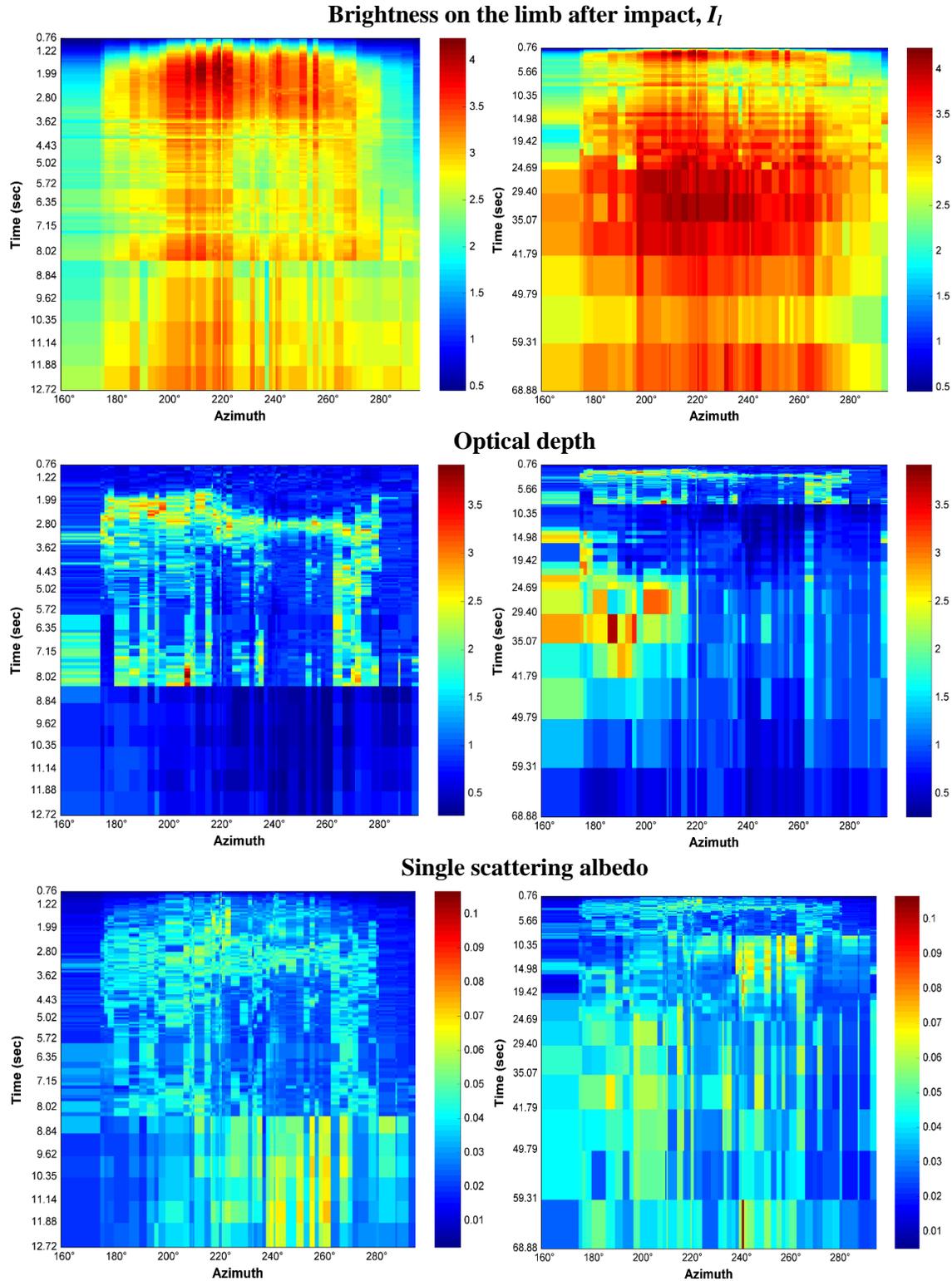

**Figure 3.** Brightness measured on the limb after impact, optical depth, and single-scattering albedo of the ejecta with the time (vertical axis) from the impact and azimuth (horizontal axis) for the first 13 seconds (left panel) and 68.8 s (right panel) after the impact. For times earlier than 0.76 s there are no reliable data on the limb obscuration.

### 3. Discussion: Crater development and depth of the excavated material during the early stages after impact.

To understand the nature of the features seen in the images shown in Figure 3, we need to understand from what depth and what location on the surface they originated. This required modeling the crater formation process, which is presented in this section.

The most important thing we need to define first is whether the crater formed under gravity or strength domination. This determines the key parameters of crater formation, and specifically the duration of the formation, the mass of the excavated material and the formulae we use for the analysis of the ejecta excavation depth.

By analyzing the post-impact crater seen in Stardust-NExT images, Schultz et al. (2013) conclude that either the crater was entirely gravity dominated or that a nested crater was formed, with the primary portion (which is the subject of this study), responsible for the crater diameter, being gravity dominated. Richardson and Melosh (2006) also supported the gravity domination; they wrote that "The resulting solid ejecta plume, highly visible due to its extremely small particle distribution, displayed classic gravity-dominated cratering behavior during its ejection and observed fallback phases: forming an inverted, conical cloud of launched particles which remained attached to the comet's surface and slowly expanded over the course of the observations." Although later Richardson and Melosh (2013), analyzing the same Stardust NExT images, as Schultz et al. conclude that the crater was strength dominated, we find this in contradiction not only with the other estimates of the crater formation scenario, but also with the estimates of the total mass of the ejected material. Richardson and Melosh (2013) estimate the total ejecta mass resulted from a formation of strength dominated crater between $5.4*10^5$ kg and $2.6*10^6$ kg. However, this is lower than the mass of the Deep Impact ejecta. Keller et al. (2007), using observations by OSIRIS instrument onboard the Rosetta spacecraft, estimated that only water molecules, without other ejecta components, constituted $(4.5-9) \times 10^6$ kg and the total mass of the ejecta should be between $10^7$ kg and $10^8$ kg. Similar numbers were obtained from ground-based observations by Schleicher et al. (2006) and Biver et al. (2007). Our estimate (Nagdimunov et al. 2014), possible only for a gravity controlled crater, was $3*10^7$ kg. Notice also that a strength dominated crater should be formed quite quickly, within several seconds, whereas a gravity dominated crater would require hundreds of second for its formation (Richardson et al. 2007) and the last fact is more consistent with the formation of the Deep Impact crater, specifically, in the MRI images (McLaughlin et al. 2014b) we see no evidence of detachment of the ejecta from the nucleus surface even 12 minutes after impact, which is more consistent with continuous excavation of the nucleus material for a period longer than a couple of seconds after impact.

Recently, gravity domination at the comet crater formation received additional support from the studies of the properties of comet 67P Churyumov-Gerasimenko. Groussin et al. (2015) found a low value of the strength (<15 Pa). This value is typical for meter-sized scale on the surface, which is the scale of our interest for Deep Impact crater, although on a small scale, of centimeter size, the strength can be larger, reaching ~10 kPa (Groussin et al. 2015; see also Vincent et al. 2015).

The basic equations we used are from the papers by O'Keefe and Ahrens (1993) and Holsapple and Schmidt (1987), and the parameters of the impact are from Holsapple and Housen (2007). O'Keefe and Ahrens consider three main regimes of the crater growth (Holsapple and Schmidt refer to the same 3 regimes as "very early time", "intermediate time" and "late time"):

(1) Penetration regime. It is characterized by the transfer of the projectile kinetic energy to the surface of the impacted object. The condition of penetration regime is $U*t/a<5.1$, where $U$ is the normal component of the impact velocity, $a$ is equivalent spherical radius of impactor and $t$ is time.

(2) Inertial regime. It is characterized by the expansion of the crater cavity with a constant geometry of the cavity (hemispherical). The end of inertial regime for crater depth is characterized by the time of maximum penetration, $t_{mp}$. The end of the inertial regime for crater width, which is also the time of termination of crater formation, is characterized by $t_{mw}$. For a gravity dominated crater, these can be estimated from the following formulas:

$U*t_{mp}/a = 0.92 \, (g*a/U^2)^{-(1+\mu)/(2+\mu)}$ (1)
$U*t_{mw}/a = 1.8 \, (g*a/U^2)^{-(1+\mu)/(2+\mu)}$ (2)

where $g$ is surface gravity, and $\mu$ is the coupling exponent (for details, see Holsapple and Schmidt, 1987).

(3) Terminal regime. It begins when the growth of the crater is stopped by strength and/or gravitational forces. At this time the crater lip has stopped expanding laterally and collapses into an outward propagating wave.

The formulae above were derived for gravity dominated crater formation and will be used in the following analysis. First, we estimate in which regime the crater formation was during the period of our observations.

Taking values of $U$= 5.1 km/sec and $a$=0.46 m from Holsapple and Housen (2007), we obtain that even for the earliest time of our observations, 0.5 sec, we are out of the penetration regime because $U*t/a \gg 5.1$. For the following estimates, we consider gravity dominated cases on opposite extremes presented by Holsapple and Housen: one is called "sand" and represents some porous material, and the other is called "water", it represents a non-porous material. The difference between these materials is primarily caused by different values of the coupling exponent: $\mu$=0.41 for "sand", and $\mu$=0.55 for "water". We use the results by Holsapple and Housen as they present the most recent attempts to study gravity-dominated cometary crater excavation and specifically target Deep Impact cratering, although their consideration is limited by the available experimental data on cratering, and may not fully reproduce the characteristics of cometary material. Using these values, we obtain a time of the maximum penetration, $t_{mp}$, from Eq. (1), equal to 258 s for the "sand" case and 451 s for the "water" case; thus, for both cases our time period is within the inertial regime. The time of termination of crater formation $t_{mw}$, obtained from Eq. (2), is 504 and 902 seconds, or approximately double $t_{mp}$, thus, it is even later in the crater formation period. These results show that assuming a gravity dominated excavation, the time period of the observations we use in our analysis is within the inertial period, i.e. within the period of transient crater growth (i.e. before the end of excavation and before the crater undergoes collapse or rebounding). In this case, we can use the formulas from Table 1 in Holsapple and Housen (2007) to estimate the crater depth and size as it was developing during the early stages of the crater formation. The growth of the crater diameter, $D$, is described by the formula:

$$D/a = K(\delta/\rho)^{0.4/(1+\mu)} (Ut/a)^{\mu/(1+\mu)} \qquad (3)$$

where $K$ =2.1 for the "sand" case, and $K$ = 2.6 for the "water" case. We use Eq. (3) to determine the size of the crater for specific times of observations and produce a range of values using, as before, two extreme cases: porous and non-porous materials. Note that Holsapple and Housen found that the final crater diameter should be about 88 m for porous material ("sand") and of the order of 350 m for non-porous material ("water"); thus, we may assume that the real crater with the diameter ~200 m (Schultz et al. 2013) was produced by a material whose mechanical characteristics are between these two cases. However, the difference between these two materials is unlikely to be confined solely to differing macroscopic porosity, microscopic (individual particle) porosity and other mechanical factors would also influence material behavior. To avoid the confusion that the only difference between the extreme cases is their porosity, we will refer to them primarily as "sand" and "water," using these words as nicknames, which identify the extreme cases of mechanical properties represented in Eq. (3) by different values of parameters $K$ and $\mu$. We use the formula $D/d$=1.67 from O'Keefe and Ahrens (1993) to estimate the depth of excavation, $d$, for each size of the transient crater to see from which depth the material was excavated at specific times of observations.

Although the data in Fig. 3 cover a time period up to 68.8 seconds after impact, the ejecta that crossed the limb at the observation time $t_{obs}$ were excavated at some earlier time which we call excavation time $t_{exc}$; this accounts for the travel time of the ejecta from the crater to the limb. The difference between the time when the material was excavated and the time when we observe this material crossing the limb can be estimated as,

$$t_{obs} - t_{exc} = (t_{obs\_i} - t_{exc\_i}) * v_i / v$$

where $v$ is the velocity of the material excavated at the time of excavation $t_{exc}$, and $v_i$ is the velocity of the material that crosses the limb first, i.e. its observation time is the time when we see the earliest part of the main ejecta crossing the limb (in the formula above there are projected velocities, but the projection factors cancel when we take the ratio, so we can use the unprojected velocities here; later we calculate them based on the crater-scaling law). We denote this earliest observation time as $t_{obs\_i}$, it can be taken from Fig. 3 to be 0.76 s. The corresponding excavation time, i.e., the excavation time of the material that reached the limb with the velocity $v_i$, is denoted as $t_{exc\_i}$, The value of $v$ is dependent on the excavation time, following the formula (see Hermalyn and Schultz, 2011; Richardson et al., 2007),

$$v/v_i = (t_{exc}/t_{exc\_i})^{-1/(1+\mu)}$$

where $\mu$ is the coupling exponent. Combining the two formulas above we obtain that the time after impact when we observe the material on the limb (this is the time indicated in Fig. 3) is,

$$t_{obs} = t_{exc} + (t_{obs\_i} - t_{exc\_i}) * (t_{exc}/t_{exc\_i})^{-1/(1+\mu)} \qquad (4)$$

This formula can be used to estimate the excavation time for different observation times shown in Fig. 3. This excavation time can then be used in Eq. (3) to estimate the depth and diameter of the crater at different moments after impact.

However, before we can use Eq. (4), we need to know the time of excavation of the earliest ejecta, $t_{exc\_i}$. We estimated this time with two methods using MRI images taken during the first second after impact. First, we found the image which showed the moment when the main ejecta plume just started developing (image m9000910_066, see McLaughlin et al. 2014b). Via this image we estimated the time when the excavation started, assuming it equal to the middle of the exposure time. This yielded $t_{exc\_i} \simeq 0.15$ s. We also estimated $t_{exc\_i}$ based on the discussion in A'Hearn et al (2005a), where a set of early MRI images was presented, including the image m9000910_069, which shows the moment when material from the hot flash first reached the limb. A'Hearn et al. (2005a) estimated the projected velocity of this hot material to be 5 km/s. The pixel spatial resolution (86.843 m/pixel) and number of pixels between the impact point and the limb (equal to 12) allow us to find out the projected distance from the crater to the limb, from which we can estimate the time when the hot material was ejected that is the time when the main ejecta excavation started. It was found to be $t_{exc\_i} \simeq 0.14$ s. The two estimates of $t_{exc\_i}$ are not completely identical, however, they are very close and provide extremely similar results. Since the second approach is based on several uncertain data (exact moment when the hot material reached the limb and speed of the material) whereas the uncertainty in the first approach is only related to the exact time of the first appearance of the ejected material in the image during the exposure period, in our following consideration we will focus on $t_{exc\_i} \simeq 0.15$ s.

Based on Eq. (3), the correspondence of excavation time to observation time calculated from Fig. 3 using Eq. (4), and the initial excavation time $t_{exc\_i} \simeq 0.15$ s, we have produced Figure 4 which shows the excavation depth and transient crater diameter for different moments after impact, considering the two cases described above: porous ("sand") and non-porous ("water") materials. The results for specific bands and azimuthal features are presented in Table 1, where we have also compared the computations for $t_{exc\_i} \simeq 0.15$ s and $t_{exc\_i} \simeq 0.14$ s to be sure that the uncertainty in the beginning of excavation does not affect the main conclusions of the paper.

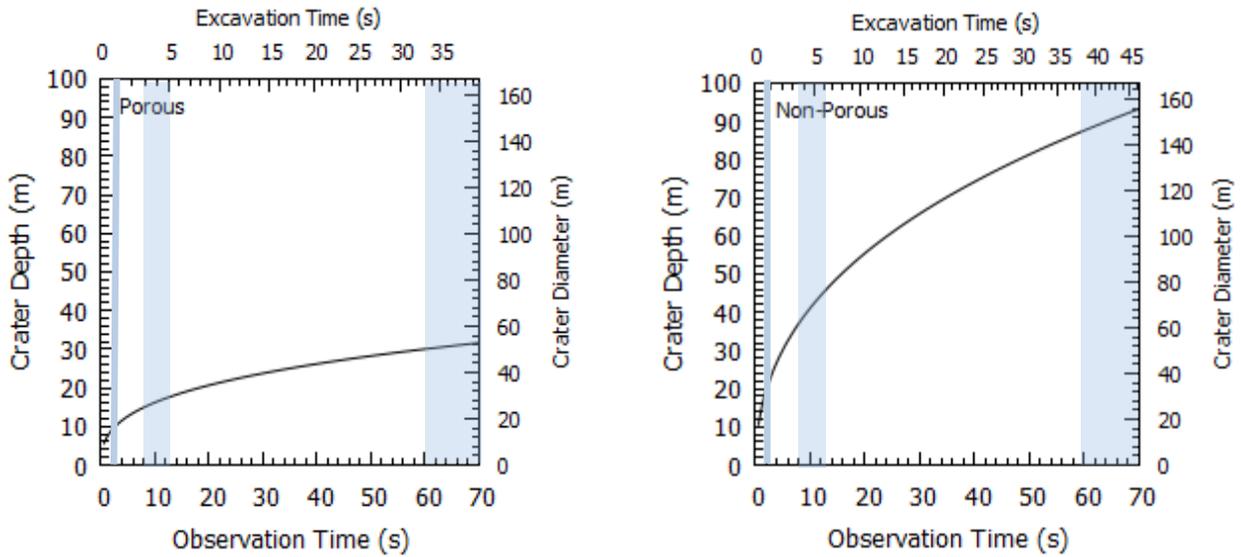

Figure 4. Calculated excavation depth (left vertical axis) and diameter (right vertical axis) of the transient Deep Impact crater as a function of observation time (lower axis) and excavation time (upper axis). The results are for a gravity dominated crater; left panel shows porous ("sand") material and the right panel shows non-porous ("water") material. The gray areas show the areas of 2-3 s, 8-13 s and 60- 68.8 s bands in optical depth seen in Fig. 3.

**Table 1. Time of excavation and depth of the excavation for porous and non-porous materials for the bands and azimuthal features in Fig. 3.**

|  | $t_{exc\_i} \simeq 0.15$ s | | | |
|---|---|---|---|---|
|  | *Porous material* | | *Non-porous material* | |
| $t_{obs}$, s | $t_{exc}$, s | depth, m | $t_{exc}$, s | depth, m |
| *Band of high optical depth* | | | | |
| 2 - 3 | 0.5 – 0.9 | 9 – 11 | 0.6 – 1 | 20 – 23 |
| *Bands of low optical depth* | | | | |
| 8 - 13 | 3 – 5.5 | 15 – 18 | 3.5 – 6 | 37 – 46 |
| 60 - 68.88 | 33 – 38 | 30 – 32 | 38.5 – 45 | 87 – 93 |
| *Azimuthal feature (high albedo) at 240° - 270°* | | | | |
| 8.5 - 20 | 3.2 – 9 | 15 – 20 | 3.7 – 10 | 38 – 55 |
| *Azimuthal feature (high optical depth) at 160° - 210°* | | | | |
| 15 - 40 | 6.4 – 20 | 18 – 26 | 7.5 – 24 | 49 – 74 |
| | | | | |

A combination of Figures 3 and 4 reveal some details about the structure of the nucleus of comet 9P/Tempel 1. Noting that the real case is likely closer to the porous ("sand") case than the non-porous ("water") case, hereafter we will cite values for the "sand" case with numbers for the "water" case in parentheses. First, as we mentioned in **Section 2.2,** there are "horizontal" (along the same time mark) inhomogeneities in the images. Particularly, there are the bands of low optical depth at 8-13 s and 60- 68.8 s of observation time. They can be interpreted as layers located on the depth 15 - 18 m (37 - 46 m) and 30 – 32 m (87 - 93 m)[1]. Low optical depth in these layers is accompanied by a higher albedo of the material, especially evident in the layer between the observation time 8 and 13 s. However, albedo appears to be more inhomogeneous than optical depth; its features are usually concentrated in a narrow range of azimuths, specifically, for the 8-13 s layer albedo is high within an azimuth of 240° - 270°. There also is a noticeable band of high optical depth located just after 2.0 seconds, i.e. slightly deeper than 9 (20) meters. This band is accompanied by an azimuthal albedo feature located at azimuths 240° - 270°. These three bands, at 2.0, 8.0, and 60 seconds are the only ones that can potentially be associated with nucleus layering. As mentioned above, the nucleus can have a finer layering structure; however, we cannot resolve it due to averaging resulting from the limited temporal and spatial resolution in the MRI data. All other features relate to azimuthal inhomogeneity. The most noticeable among them is the area of increased optical depth at 15-40 s after impact, having azimuths of 160° - 210° and a depth of about 18 – 26 (49 - 74) meters, and the already mentioned area of high albedo at 240° - 270° that extends from 15 (38) m to 20 ( 55) m. There are also narrow vertical features in the brightness which can be also seen in the optical depth and, although less evidently, in albedo. We believe these are the rays clearly seen in the images of the ejecta cloud (Fig. 1 right). These vertical features are more pronounced in the optical depth than in albedo, indicating that the rays have a characteristically higher dust density. There are also very narrow optical depth (at ~ 7 s and azimuth ~210°) and high albedo (at ~ 60 s and azimuth ~240°) features.

---

[1] We may even notice that after the 8-13 s layer the optical depth of the nucleus became generally lower (except of a chunk of high optical depth material located at azimuths 160°-210°), probably indicating increasing porosity of the nucleus. This is consistent with decrease of the dielectric constant of the nucleus materials with depth found for comet 67P/ Churyumov-Gerasimenko by CONSERT instrument on Rosetta mission (Ciarletti et al. 2015).

The resolution of crater depth and diameter based on limb crossing observations can be determined by examining individual MRI observation times, combined with our calculations summarized by Fig. 4. We obtain a crater depth resolution of as 0.2 m (for "sand") and 0.5 m (for "water") at the time the ejecta initially reached the limb, $t_{obs}$ = 0.76 s. In either of these cases, the data do not have sufficient resolution to see the thin layer of organic material reported by Kadono et al. (2007). At slightly later observation times of 7-8 s, due to the slower change in the velocities (see Fig. 4) and prior to the MRI switching to a mode with worse temporal resolution, we could obtain finer depth resolution which may be as low as 0.05 m . Note that these estimates are affected by the uncertainties in the excavation start time and rough representation of the material properties.

As we have shown in Section 2.1, after we take into account the nucleus surface inhomogeneity, the brightness of the shadow that the ejecta cloud casts on the nucleus does not show any variations which can be clearly associated with the inhomogeneity of the ejecta cloud. Unfortunately, analysis of the shadow ejecta is much more complicated than the analysis of the limb-crossing ejecta. The observations of the limb-crossing ejecta show mainly the dust that forms the leading edge of the ejecta cone, i.e. the ejecta which move in the direction opposite to the impactor trajectory (see Fig. 3 in Schultz et al. 2007) whereas the shadow is formed by the light that goes through the whole ejecta plume. Specifically, the shadow is partly formed by the light that passes through the ejecta in the central part of the cone, which are likely extracted from different layers than the outer parts of the ejecta cone (Schultz et al. 2007). Besides, the outer parts of the ejecta cone are affected by azimuthal difference in the ejecta speed that is typical for oblique impacts (see Anderson et al. 2003 and Richardson 2011). The latter results in azimuthal variations of the calculated ejecta excavation time, and, thus, in azimuthal variations of their calculated excavation depth. Since in the shadow, we observe material from different depths simultaneously, determining the excavation time of the ejecta which cast each part of the shadow is very problematic. We suppose that during the shadow observation (0.96 – 2 s after impact) we see the material excavated from the depth from 9 m to 15 m for "sand" (or to 30 m for "water"), but cannot disentangle depth or excavation time for individual pixels.

4. **Conclusions**

In the paper we presented two approaches which allow one to characterize the materials excavated during an impact crater formation at different moments after impact. Both approaches were tested using the images of the ejecta resulted from the Deep Impact experiment on comet 9P Tempel 1.

The most interesting result of our analysis is that observed inhomogeneities (bands and azimuthal features in albedo or optical depth) in the ejecta imply inhomogeneities both laterally and with depth in the tens of meters near the surface of the nucleus. We then use crater-scaling-laws to estimate from which depth the material that produces said albedo or optical depth inhomogeneities was excavated, and use this as a probe of the internal inhomogeneity of the comet nucleus. Summarized characteristics of these bands and azimuthal features are presented in Table 1. It is hard to define the physical reasons that are responsible for the described bands and azimuthal features. Whereas high albedo may be associated with a higher abundance of ice and low albedo with a low ice abundance or domination of absorbing (perhaps carbonaceous) materials, the inhomogeneities in the optical depth are harder to explain unambiguously. Our radiative transfer modeling reported in Nagdimunov et al. (2014) showed that, due to the fact that the optical depth is a function of the extinction cross-section multiplied by the number density, we can find a variety of combinations of extinction cross-section and number density that produce the same fit for the optical depth. The bands of low optical depth or azimuthal inhomogeneities in optical depth can be areas of lower bulk density of the material (due to higher porosity) or areas characterized by dust consisting of particles of smaller extinction cross-section. If there is no albedo feature correlated with the optical depth feature, then, more likely, the reason of the optical depth feature is a varying porosity (number density), although variations in the particle size are also possible. If there is a correlation between the optical depth and albedo variations, then, more likely, a difference in the dust particle extinction is the reason of the variations, indicating either a difference in material absorption, or in particle size variations, or both.

**Acknowledgement**

The work was supported by the NASA PMDAP program, grant NNX10AP31G. We appreciate comments by the anonymous reviewer which led to a significant improvement of the paper.

**References**
A'Hearn, M, M. Belton, W. A. Delamere, J. Kissel, K. P. Klaasen, L. A. McFadden, et al., 2005a, Deep Impact: Excavating Comet Tempel 1," Science 310, 258-264.
A'Hearn, M., M. Belton, A. Delamere, W. H. Blume, 2005b, Deep Impact: A large-scale active experiment on a cometary nucleus, Space Sci. Rev. 117, 1-21.
Anderson, J. L., Schultz, P. H., and Heineck, J. T., 2003, Asymmetry of ejecta flow during oblique impacts using three-dimensional particle image velocimetry. Journal of Geophysical Research: Planets, 108(E8), 13-1 – 13-10.
Belton, M. J., Thomas, P., Veverka, J., Schultz, P., A'Hearn, M. F., Feaga, L., ... and Kissel, J., 2007, The internal structure of Jupiter family cometary nuclei from Deep Impact observations: The "talps" or "layered pile" model. *Icarus*, *187*(1), 332-344.
Biver, N. D., Bockelée-Morvan, D., Boissier, J., Crovisier, J., Colom, P., Lecacheux, A., Moreno, R., Paubert, G., Lis, D. C., Sumner, M. et al., 2007, Radio observations of comet 9P/Tempel 1 before and after deep impact. Icarus, 187, 253–271.
Ciarletti, V. A.C. Levasseur-Regourd, J. Lasue, C. Statz, D. Plettemeier, A. Hérique, Y. Rogez and W. Kofman, 2015, CONSERT suggests a change in local properties of 67P/Churyumov-Gerasimenko's nucleus at depth, Aastron.Astrophys., DOI: http://dx.doi.org/10.1051/0004-6361/201526337.

Hampton, D. L., Baer, J. W., Huisjen, M. A., Varner, C. C., Delamere, A., Wellnitz, D. D., ... and Klaasen, K. P., 2005, An overview of the instrument suite for the Deep Impact mission. *Space Science Reviews*, *117*(1-2), 43-93.

Hermalyn, B., and Schultz, P. H. 2011. Time-resolved studies of hypervelocity vertical impacts into porous particulate targets: Effects of projectile density on early-time coupling and crater growth. Icarus, 216(1), 269-279.

Holsapple, K. A., and Schmidt, R. M., 1987. Point source solutions and coupling parameters in cratering mechanics. Journal of Geophysical Research: Solid Earth (1978–2012), 92(B7), 6350-6376.

Holsapple, K.A., Housen, K.R., 2007. A crater and its ejecta: An interpretation of Deep Impact. Icarus 187, 345–356.

Groussin, O., L. Jorda, A.-T. Auger, E. Kührt, R. Gaskell, C. Capanna, F. Scholten, F. Preusker, P. Lamy, S. Hviid, J. Knollenberg, U. Keller, C. Huettig, H. Sierks, C. Barbieri, R. Rodrigo, D. Koschny, H. Rickman, M. F. A'Hearn, J. Agarwal, M. A. Barucci, J.-L. Bertaux, I. Bertini, S. Boudreault, G. Cremonese, V. Da Deppo, B. Davidsson, S. Debei, M. De Cecco19, M. R. El-Maarry, S. Fornasier, M. Fulle, P. J. Gutiérrez, C. Güttler, W.-H Ip, J.-R. Kramm, M. Küppers, M. Lazzarin, L. M. Lara, J. J. Lopez Moreno, S. Marchi, F. Marzari, M. Massironi, H. Michalik, G. Naletto, N. Oklay, A. Pommerol, M. Pajola, N. Thomas, I. Toth, C. Tubiana, and J.-B. Vincent, 2015, Gravitational slopes, geomorphology, and material strengths of the nucleus of comet 67P/Churyumov-Gerasimenko from OSIRIS observations Astron. and Astrophys. manuscript no. 26379_ap cESO 2015 July 21, 2015

Kadono, T., Sugita, S., Sako, S., Ootsubo, T., Honda, M., Kawakita, H., Miyata, T., Furusho, R., and Watanabe, J., 2007. The thickness and formation age of the surface layer on comet 9P/Tempel 1. Ap.J.Let. 661(1), L89 – L92.

Keller, H.U., Küppers, M., Fornasier, S., Gutierrez, P.J., Hviid, S.F., Jorda, L., Knollenberg, J., Lowry, S.C., Rengel, M., Bertini, I., Cremonese, G., Ip, W.-H., Koschny, D., Kramm, R., Kührt, E., Lara, L.-M., Sierks, H., Thomas, N., Barbieri, C., Lamy, Ph., Rickman, H., Rodrigo, R., A´Hearn, M.F., Angrilli, F., Barucci, M.-A.,. Bertaux, J.-L, da Deppo, V., Davidsson, B. J. R., de Cecco, M., Debei, S., Fulle, M., Gliem, F., Groussin, O., Lopez Moreno, J.J., Marzari, F., Naletto, G., Sabau, L., Andrés, S. A., Wenzel, K.-P., 2007. Observations of Comet 9P/Tempel 1 around the Deep Impact event by the OSIRIS cameras onboard Rosetta. Icarus 191, 241-257.

Kokhanovsky, A., 2004, Light scattering media optics: problems and solutions, Springer, Berlin, 240 p.

McLaughlin, S. A., B. Carcich, T. McCarthy, M. Desnoyer, and K.P. Klaasen, 2014a, DEEP IMPACT 9P/TEMPEL ENCOUNTER - REDUCED HRIV IMAGES V3.0, DIF-C-HRIV-3/4-9P-ENCOUNTER-V3.0, NASA Planetary Data System.

McLaughlin, S.A., B. Carcich, S.E. Sackett, T. McCarthy, M. Desnoyer, K.P. Klaasen, and D.W. Wellnitz, 2014b. DEEP IMPACT 9P/TEMPEL ENCOUNTER - REDUCED MRI IMAGES V3.0, DIF-C-MRI-3/4-9P-ENCOUNTER-V3.0, NASA Planetary Data System

Meech, K. J., Ageorges, N., A'Hearn, M. F., Arpigny, C., Ates, A., Aycock, J.... and Filipovic, M. D., 2005, Deep Impact: Observations from a worldwide Earth-based campaign, Science, 310(5746), 265-269.

Nagdimunov, L., Kolokolova, L., Wolff, M., A'Hearn, M. F., and Farnham, T. L., 2014, Properties of comet 9P/Tempel 1 dust immediately following excavation by Deep Impact. Planetary and Space Science, 100, 73-78.

O'Keefe, J. D., and Ahrens, T. J. 1993. Planetary cratering mechanics. Journal of Geophysical Research: Planets (1991–2012), 98(E9), 17011-17028.

Richardson, J. E., 2011, Modeling impact ejecta plume evolution: A comparison to laboratory studies. Journal of Geophysical Research: Planets, 116(E12), CiteID E12004.

Richardson, J. E.; Melosh, H. J.(2006) Modeling the Ballistic Behavior of Solid Ejecta from the Deep Impact Cratering Event, 37th Annual Lunar and Planetary Science Conference, March 13-17, 2006, League City, Texas, abstract no.1836.

Richardson, J.E., Melosh, H.J., Lisse, C.M., Carcich, B, 2007, A ballistics analysis of the Deep Impact ejecta plume: Determining Comet Tempel 1's gravity, mass, and density. Icarus 191, 176-209.

Richardson, J. E., and Melosh, H. J., 2013. An examination of the Deep Impact collision site on Comet Tempel 1 via Stardust-NExT: Placing further constraints on cometary surface properties. Icarus, 222(2), 492-501.

Robitaille, T.P., 2011, HYPERION: an open-source parallelized three-dimensional dust continuum radiative transfer code. Astron. Astrophys. 536, A79.

Schleicher, D. G., K. L. Barnes, N. F. Baugh, 2006. Photometry and imaging results for comet 9P/Tempel 1 and Deep Impact: Gas production rates, postimpact light curves, and ejecta plume morphology. Astron. J. 131, 1130–1137.

Schultz, P.H., Eberhardy, C.A., Ernst, C.M., A'Hearn, M.F., Sunshine, J.M., Lisse, C.M., 2007, The Deep Impact oblique impact cratering experiment. Icarus, 191, 84-122.

Schultz, P. H., Hermalyn, B., and Veverka, J., 2013. The Deep Impact crater on 9P/Tempel-1 from Stardust-NExT. Icarus, 222(2), 502-515.

Vincent, J. B., Oklay, N., Marchi, S., Höfner, S., and Sierks, H., 2015, Craters on comets. Planetary and Space Science, 107, 53-63.